\documentclass[RNAAS]{aastex62}

\graphicspath{{./}{figures/}}

\begin{document}

\title{Efficient joint sampling of impact parameters and transit depths in transiting exoplanet lightcurves}

\correspondingauthor{N\'estor Espinoza}
\email{espinoza@mpia.de}

\author[0000-0001-9513-1449]{N\'estor Espinoza}
\altaffiliation{Bernoulli Fellow, IAU-Gruber Fellow}
\affiliation{Max-Planck-Institut f\"ur Astronomie, K\"onigstuhl 17, 69117 Heidelberg, Germany.}


\section{Introduction} \label{sec:intro}

When fitting transiting exoplanet lightcurves, it is usually desirable to have ranges and/or priors for the parameters which are to be retrieved that include our degree of knowledge (or ignorance) in the routines which are being used. In Markov Chain Monte Carlo (MCMC) routines, for example, these enter as \citep[hopefully proper;][]{Tak:2018} prior distributions. These can either represent our current knowledge of the distribution of such parameters \citep[e.g., based on their observed values, see, e.g., ][]{Kipping:2013-ecc, Kipping:2014} or physically plausible parameters ranges to be sampled \citep[see, e.g., ][]{Kipping:2013, Kipping:2016}. 

Among the parameters that are constrained by transiting exoplanet lightcurves, there are two which are of much physical significance: the impact parameter of the orbit, $b = (a/R_*)\cos i $, and the planet-to-star radius ratio, $p = R_p/R_s$ (which defines the transit depth, $\delta = p^2$). These two are natural parameters to extract and constrain as they usually have well defined limits: $b$ is only physically meaningful on a transiting system for values $0< b < 1 + p$ while $p$ can usually be defined for values $0<p<1$ \citep[except, e.g., when looking for planets around white dwarfs; see ][]{Agol:2011}. A common set of ``uninformative'' priors used for those two parameters are uniform priors. On one hand, the prior on $p$ can be taken to be uniform over a lower and upper limit, $p_l$ and $p_u$, while $b$ can be taken to be uniform between its maximum possible range given the prior on $p$, $(0, 1 + p_u)$. However, this poses a sampling problem especially important for grazing orbits. Given that we sample a value $p_i$ from the prior on $p$, the only physically plausible values for $b$ to be sampled given $p_i$ are those that satisfy $b < 1 + p_i$ (see Figure \ref{fig:1}, left panel; black points). If we simply reject the sample if the sampled value of b is greater than $1 + p_i$ (grey points in Figure \ref{fig:1}), then we will reject points from a significant portion of the prior area (to the right of the blue dashed line in Figure \ref{fig:1}; covering there 25\% of the prior area). It is desirable, thus, to have an algorithm that efficiently samples values from the physically plausible zone in the $(b,p)$ plane, which is the quadrilateral $ABED$ sampled by the black points in Figure \ref{fig:1}. Here we present such an algorithm.

\section{The algorithm}
\begin{figure}[h!]
\begin{center}
\includegraphics[scale=0.43,angle=0]{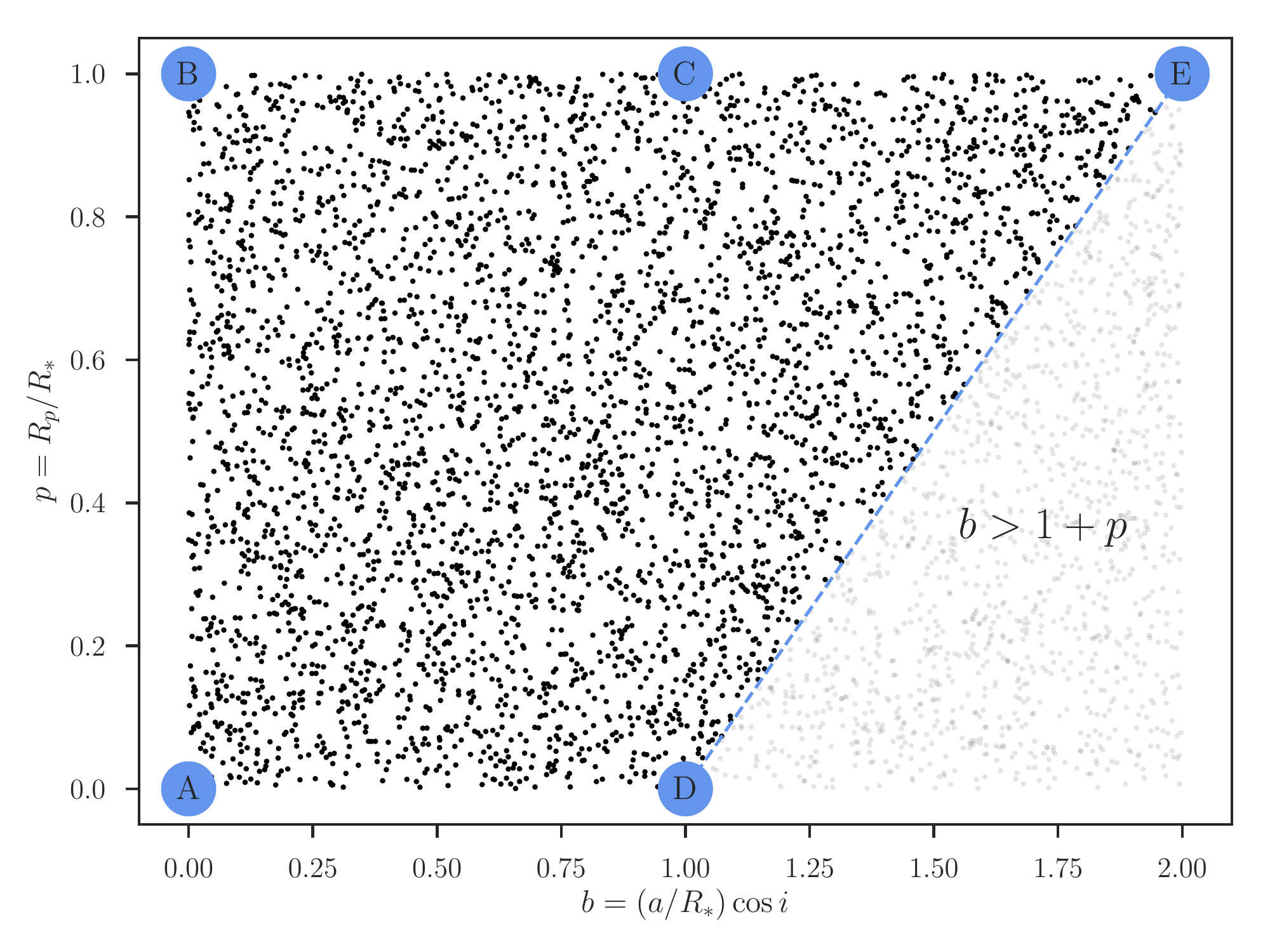}
\includegraphics[scale=0.43,angle=0]{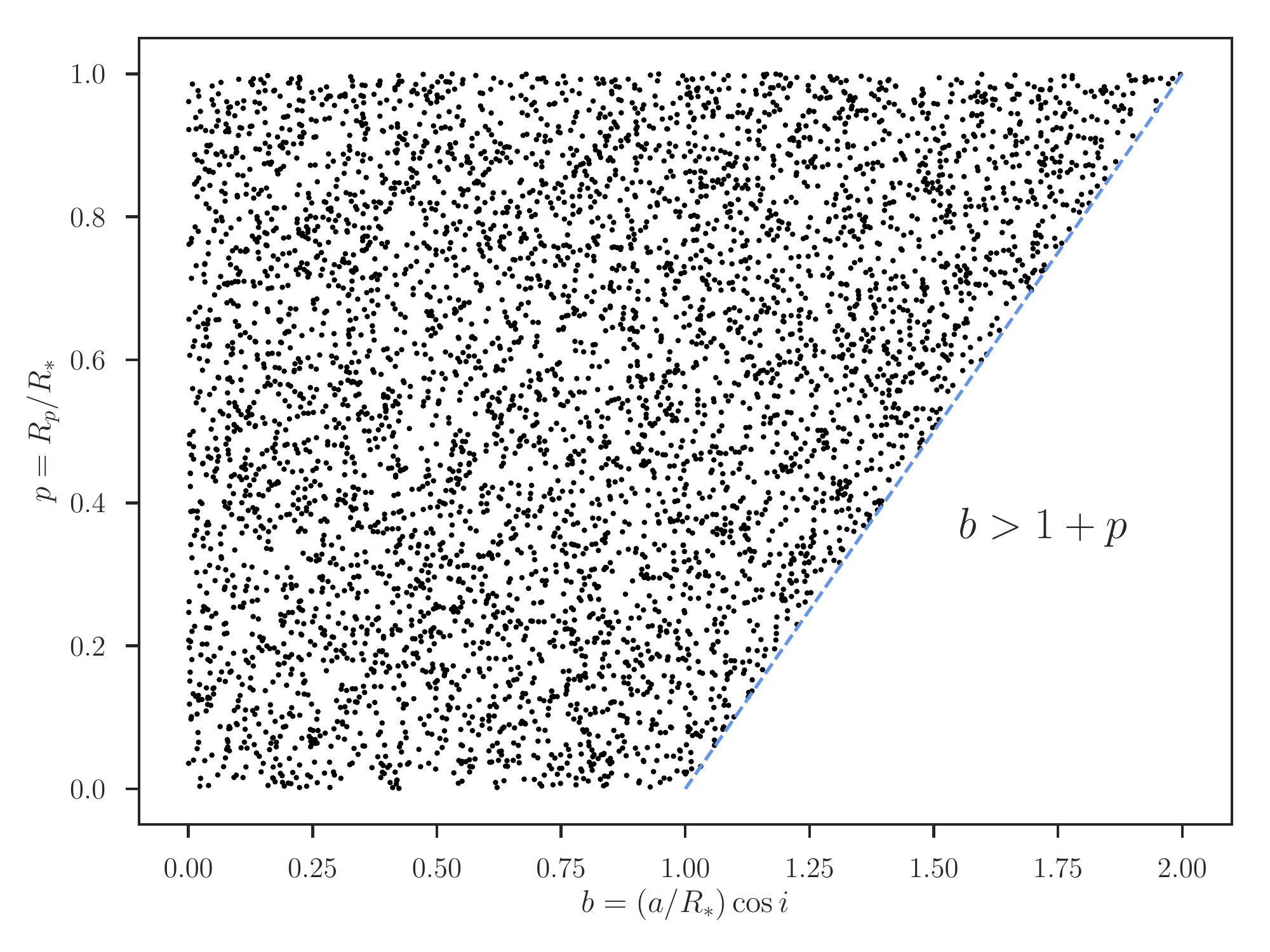}
\caption{(Left) Samples generated for $b$ and $p$ using rejection sampling (black and grey points) for the case $(p_u,p_l)=(0,1)$. Grey points denote the unphysical regime for the $(b,p)$ pairs (see text), with the dashed blue line indicating the limit at which this happens ($b=1+p$). Blue dots with letters indicate the vertices of the geometry of the problem outlined in the text (Right) Samples generated with the algorithm outlined in this Research Note.\label{fig:1}}
\end{center}
\end{figure}

The basic idea of the algorithm is quite simple if one notes that the quadrilateral $ABED$ in Figure \ref{fig:1} can be decomposed into a rectangle (with vertices in the $(b,p)$ plane at 
$A = (0,p_l)$, $B = (0,p_u)$, $C=(1+p_l,p_u)$ and $D = (1+p_l,p_l)$) and a triangle 
(with vertices $C$, $D$ and $E = (1+p_u, p_u)$). Given this decomposition, the idea is to first select either with probability equal \textit{to the fraction of area covered by them} ($A_r = A_{CDE}/(A_{CDE} + A_{ABCD})$ for the triangle and $1-A_r$ for the rectangle), sample 
random points inside the chosen geometric shape, and then repeat the process. Sampling uniform points in the $ABCD$ rectangle is straightforward. To sample points from the $CDE$ triangle, one can use the triangular sampling technique outlined in \cite{Turk:1990}, and already used by \cite{Kipping:2013} in the context of sampling physically meaningful limb-darkening coefficients. With this idea in mind, the following algorithm requiring two random numbers $r_1\sim U(0,1)$ and $r_2\sim U(0,1)$, where $U(a,b)$ stands for a uniform distribution between $a$ and $b$, along with the value of $A_r = (p_u-p_l)/(2+p_l+p_u)$, 
can efficiently sample points from the quadrilateral $ABED$: 
\begin{enumerate}
\item If $r_1 > A_r$ go to step 2. If $r_1 \leq A_r$, go to step 3.
\item In this case we want to sample points uniformly inside the $ABCD$ rectangle. 
To this end, we note that if we are in this step, then $r_1\sim U(A_r,1)$. Sample then a $(b,p)$ 
pair so that $b,p\in ABCD$ using the transformations
\begin{eqnarray}
b &=& [1+p_l][1+(r_1-1)/(1-A_r)]\\
p &=& (1-r_2)p_l + r_2p_u.
\end{eqnarray}
This will generate $(b,p)$ random variates with $b \sim U(0,1+p_l)$ and $p\sim U(p_l,p_u)$.
\item In this case we want to sample points from the CDE triangle. For this we use the technique outlined in \cite{Turk:1990} and pedagogically explained in \cite{Kipping:2013}, in which a point $b,p \in CDE$ 
can be sampled given two random variates 
$q_1\sim U(0,1)$ and $q_2\sim U(0,1)$. We already have one such random variate from 
our sampling of $r_2$; with this we have $q_2$. To get a random variate with 
the properties of $q_1$, we again note that if we are in this step, $r_1 \sim U(0,A_r)$. 
The transformation $q_1 = r_1/A_r$ makes 
$q_1\sim U(0,1)$. With this, the transformations
\begin{eqnarray}
b &=& (1+p_l) + \sqrt{q_1}q_2(p_u-p_l) = (1+p_l) + \sqrt{r_1/A_r}r_2(p_u-p_l),\\
p &=& p_u + (p_l-p_u)\sqrt{q_1}(1-q_2) = p_u + (p_l-p_u)\sqrt{r_1/A_r}(1-r_2),
\end{eqnarray}
sample points from the CDE triangle.
\end{enumerate}

Points sampled with this new technique are presented in Figure \ref{fig:1}, right panel: they sample the quadrilateral uniformly, as expected. Codes to reproduce those figures can be found at \url{https://github.com/nespinoza/impact-radius}. With this algorithm, in a transit fitting routine through, e.g., an MCMC, one can define the lower and upper limits of $p$, $p_l$ and $p_u$, and simply use $r_1$ and $r_2$ as the parameters to be fitted instead of $p$ and $b$. Once finished, the posterior distributions of $r_1$ and $r_2$ can be easily inverted using equations (1)---(4) in order to retrieve the posterior distribution of $b$ and $p$: equations (1) and (2) for all $r_1>A_r$ and equations (3) and (4) for all $r_1 \leq A_r$. We do warn, however, that sometimes this \textit{could} lead to bimodal distributions in the ($r_1$,$r_2$) plane --- sampling algorithms thus should be able to cope with this possibility.

\bibliography{bibfile}

\end{document}